\documentclass[11pt]{article}
\pdfoutput=1 

\usepackage{jheppub} 

\usepackage{amssymb,amsmath,amsthm,graphicx}
\usepackage{latexsym}
\usepackage{bm}
\usepackage[]{hyperref}
\usepackage{cleveref}
\usepackage{subcaption}
\usepackage{float}

\def \be {\begin{equation}}
\def \ee {\end{equation}}
\def \bea {\begin{eqnarray}}
\def \eea {\end{eqnarray}}

\pdfoutput=1

\begin{document}

\title{Weak cosmic censorship with excited scalar fields and bound on charge-to-mass ratio}
\author[\sharp]{Si-Yuan Cui and Yong-Qiang Wang \footnote{yqwang@lzu.edu.cn, corresponding author}}

\affiliation{$^{a}$Lanzhou Center for Theoretical Physics, Key Laboratory of Theoretical Physics of Gansu Province,
	School of Physical Science and Technology, Lanzhou University, Lanzhou 730000, People's Republic of China\\
	$^{b}$Institute of Theoretical Physics $\&$ Research Center of Gravitation, Lanzhou University, Lanzhou 730000, People's Republic of China}

\abstract{Recent study in \cite{Crisford:2017gsb}  discovered that introducing a massive charged scalar field and requiring the Weak Gravity Conjecture (WGC) to hold can eliminate a class of Weak Cosmic Censorship Conjecture (WCCC) counterexamples in anti-de Sitter spacetime, indicating a potential connection between WCCC and WGC. In
this paper, we extend the study  to the case of excited-state scalar fields, and numerically construct the static solutions of excited massive charged scalar fields coupled to the Einstein-Maxwell field in four dimensional spacetime with asymptotically anti-de Sitter boundary conditions. In the absence of scalar field, there is a class of 
counterexamples to cosmic censorship.
However, after adding the scalar field with sufficiently large charge, the original
counterexamples can be removed and cosmic censorship is preserved. We find this conclusion also applies to the excited-state of scalar field. There is a minimum value $q_c^{bound}$. When the charge of the excited scalar field is larger than this minimum value, for sufficiently large boundary electric amplitude $a$, there will not appear a region with arbitrarily large curvature. That means there exists lower bound on the charge for excited state fields which protects cosmic censorship from being violated.}

\maketitle

\section{Introductions}\label{sec1}

  Recent years have witnessed exhilarating astronomical observations related to black holes \cite{EventHorizonTelescope:2019dse,EventHorizonTelescope:2022wkp,LIGOScientific:2016aoc,LIGOScientific:2018mvr,LIGOScientific:2020ibl,KAGRA:2021vkt}, yet our understanding of one of the profound theoretical challenges about black holes - the singularity problem - remains limited. In 1969, R. Penrose proposed a conjecture related to this issue: Does there exist a “cosmic censor” who forbids the appearance of naked singularities, clothing each in an absolute event horizon? \cite{Penrose:1969pc}. This idea, was later developed and refined as the weak cosmic censorship conjecture (WCCC) \cite{Wald:1997wa}. To check this conjecture, numerous works have been done over the decades based on Wald’s gedanken experiment \cite{Wald:1974hkz}. Some of these have indeed demonstrated the conjecture's validity under certain conditions \cite{Wald:1974hkz,Hod:2008zza,Sorce:2017dst,Rocha:2011wp,Jiang:2019ige,Chen:2019nhv,Yang:2020iat,Yang:2020czk,Liang:2018wzd,Qu:2020nac}. However, high-dimensional black hole solutions with non-linear instabilities have yielded contradictory results \cite{Lehner:2010pn,Figueras:2015hkb,Figueras:2017zwa,Gregory:1993vy,Hubeny:2002xn,Santos:2015iua}. They can evolve from smooth initial data into naked singularities. 
Another particularly noteworthy class of counterexamples are solutions in four-dimensional with asymptotically anti-de Sitter (AdS) boundary conditions, where Einstein-Maxwell equations yield a region of infinite curvature observable by distant observers \cite{Horowitz:2016ezu,Crisford:2017zpi}, contravening the WCCC.

Subsequently, G.T. Horowitz and others discovered that these counterexamples to the WCCC could be eliminated by introducing a charged scalar field, suggesting a potential connection between the WCCC and the weak gravity conjecture (WGC) \cite{Crisford:2017gsb}. The introduced charged scalar field destabilizes the original Einstein-Maxwell solutions, forming new solutions with scalar `hair'. If the scalar field's charge is sufficiently large, exceeding a minimum charge $q_{min}$, regions of infinite curvature no longer appear, thereby preserving the WCCC. The weak gravity conjecture \cite{ArkaniHamed:2006dz,Palti:2019pca,Cheung:2014vva,Harlow:2022ich,Heidenreich:2017sim}, a hypothesis about the relative strength of gravitational force in quantum gravity theories, posits that in any consistent theory of quantum gravity, gravity must be the weakest force. A direct consequence of this is the requirement for the existence of a stable charged elementary particle whose charge-to-mass ratio $q / m>1$ (natural units). 
Most intriguingly, the $q_{min}$ required to preserve WCCC  coincides exactly with the lower bound for particle charge proposed by the WGC. Furthermore, this connection persists even when additional electric fields are introduced or the charged scalar field is replaced with a dilaton field \cite{Horowitz:2019eum}. In our previous work, we also calculated solutions taking into account the self-interacting scalar fields \cite{Song:2020onc}, Born-Infeld electrodynamics \cite{Hu:2019rpw}, and SU(2) fields \cite{Song:2022dpw}, similarly identifying a lower bound for field charge to maintain the WCCC.

However, all the aforementioned works focus on ground state solutions, where the added scalar or gauge fields remain in their ground state (field functions keep sign in the domain of definition). Naturally, we are also curious to explore the existence of excited state solutions (field functions change sign and have nodes along the radial direction) and verify whether a similar charge lower bound exists to protect the WCCC in these states. Actually, the study of the excited states of fields coupled to the gravitational field is a research direction that has garnered attention. For instance, there are many studies on excited state boson stars \cite{Bernal:2009zy,Collodel:2017biu,Alcubierre:2018ahf,Liebling:2012fv,Li:2019mlk,Sun:2022duv,Liang:2023ywv,Yue:2023bae,Zeng:2023hvq,Huang:2023glq}, and a recent work even found that excited state boson stars can be stable when considering self-interactions of the scalar field \cite{Sanchis-Gual:2021phr}. Research in the field of holographic superconductors has also examined excited state solutions \cite{Wang:2019caf,Wang:2019vaq,Qiao:2020fiv,Xiang:2020ugu}, as well as studies on the properties of excited state hairy black holes \cite{Wang:2018xhw,Delgado:2019prc,Kunz:2019bhm,Teodoro:2021ezj,Herdeiro:2023roz}.

In this work, we extend the study of \cite{Crisford:2017gsb} to the case of excited-state scalar fields. We construct solutions for a massive charged scalar field in excited states coupled with the Einstein-Maxwell field in four-dimensional spacetime with asymptotically anti-de Sitter boundary conditions. We present the distribution of scalar fields in the first and second excited states, and arrive at a conclusion similar to \cite{Crisford:2017gsb}: scalar fields in the first and second excited states also have a minimum charge requirement to safeguard the WCCC. It's worth noting that coupling anti-de Sitter gravity with Maxwell and charged scalar fields can produce the gravitational dual of superconductors \cite{Hartnoll:2008vx,Hartnoll:2008kx} (known as s-wave holographic superconductor \cite{Cai:2015cya}).

The structure of this paper is as follows. In Sec.2, we review the Einstein-Maxwell-scalar model in four dimensions with asymptotically anti-de Sitter boundary conditions. We also introduce our numerical calculation methods and the ansatzs and boundary conditions used in our work. In Sec.3, we present the numerical results for the scalar field in excited states, both as linear solutions (considering the scalar field as a test field against a fixed spacetime background) and non-linear solutions (where the scalar field influences the spacetime background metric), and study their properties. In the final section, we conclude our results.

.

\section{The setup}\label{sec2}
\subsection{Action and equations of motion}
  
  We consider the model in four dimensional Einstein-Maxwell-scalar theory with a negative cosmological constant. The bulk action reads
\begin{subequations}
\begin{align}
S=\frac{1}{16 \pi G} \int d^4 x \sqrt{-g}\left[R-2\Lambda-F^{a b} F_{a b}+\mathcal{L}_{\text {S}}\right],
 \label{eq:action}
\end{align}
where $G$ is the gravitational constant and $\Lambda=-\frac{3}{L^2}$ represents the cosmological constant in which $L$ is the AdS length scale, in this work we set $L=1$. $F_{a b}$ is the electromagnetic strength tensor with respect to $A$, $\mathcal{L}_{\text {S}}$ which represents the scalar field contribution reads
\begin{align}
\mathcal{L}_{\text {S}}=-4\left(\mathcal{D}_a \Phi\right)\left(\mathcal{D}^a \Phi\right)^{\dagger}-4 m^2 \Phi \Phi^{\dagger},
\end{align}
where $\mathcal{D}=\nabla-i q A$ is the covariant derivative with respect to $A$. The constants $m$ and $q$ represent the mass and the charge of the complex scalar field. 
\label{eqs:action}
\end{subequations}
The equations of motion that derived from the action (\ref{eqs:action}), can be written as
\begin{subequations}
\begin{align}
\label{eq:Einsteineq}
R_{a b}+\frac{3}{L^2} g_{a b}=2\left(F_a{ }^c F_{b c}-\frac{g_{a b}}{4} F_{c d} F^{c d}\right)+2\left(\mathcal{D}_a \Phi\right)\left(\mathcal{D}_b \Phi\right)^{\dagger}+2\left(\mathcal{D}_a \Phi\right)^{\dagger}\left(\mathcal{D}_b \Phi\right)+2 m^2 g_{a b} \Phi \Phi^{\dagger}\, ,
\end{align}
\begin{align}
\label{eq:Maxwllseq}
\nabla_a F^a{ }_b=i q\left[\left(\mathcal{D}_b \Phi\right) \Phi^{\dagger}-\left(\mathcal{D}_b \Phi\right)^{\dagger} \Phi\right] ,
\end{align}
\begin{align}
\label{eq:KGseq}
\mathcal{D}_a \mathcal{D}^a \Phi=m^2 \Phi ,
\end{align}
\label{eqs:EOM}
\end{subequations}

From the equations (\ref{eqs:EOM}), it is easily apparent that in the absence of a scalar field ($\Phi=0$), the equations reduce to the Einstein-Maxwell equations with a negative cosmological constant. In this scenario, we choose the asymptotically flat boundary metric as: 
\begin{equation}
\mathrm{d} s_{\partial}^2=-\mathrm{d} t^2+\mathrm{d} r^2+r^2 \mathrm{~d} \phi^2,
\end{equation}
and set the boundary electric potential to be
\begin{equation}
\left.A\right|_{\partial}=a(t) p(r) \mathrm{d} t,
\end{equation}
where $p(r)$ is a radial profile that diminishes at large radius faster than $1 / r$ and $a(t)$ is an amplitude. Assuming $a(t)$ changes sufficiently slowly, this process can be regarded as a combination of a series of static solutions with specified values of $a$. As the value of $a$ gradually increases to a maximum amplitude $a = a_{max}$ , a naked singularity emerges, thus providing a counterexample to the WCCC \cite{Horowitz:2016ezu}. Subsequent work \cite{Crisford:2017zpi} presents the complete nonlinear evolution process and demonstrates that regions of infinite curvature can indeed be observed by a distant observer.

\subsection{The numerical method}

  Given the difficulty of obtaining analytical solutions for the Einstein-Maxwell-scalar theory in AdS spacetime, we turn to get numerical solutions of the equations of motion (\ref{eqs:EOM}). We are aware that general relativity is invariant under coordinate transformations, therefore, (\ref{eq:Einsteineq}) can not turn into well-defined elliptic equations. Utilizing the DeTurck method \cite{Headrick:2009pv,Adam:2011dn}, we introduce a DeTurck term into the original Einstein equations (\ref{eq:Einsteineq}), transforming the complicated hyperbolic equations into solvable elliptic equations (more detail in \cite{Dias:2015nua,Wiseman:2011by}), known as the Einstein-DeTurck equations

\begin{equation}
\begin{split}
\label{eq:EinsteinDeTurck}
R_{a b}+\frac{3}{L^2} g_{a b}-\nabla_{(a} \xi_{b)}=&2\left(F_a{ }^c F_{b c}-\frac{g_{a b}}{4} F_{a d} F^{c d}\right)\\
&\quad+2\left(\mathcal{D}_a \Phi\right)\left(\mathcal{D}_b \Phi\right)^{\dagger}+2\left(\mathcal{D}_a \Phi\right)^{\dagger}\left(\mathcal{D}_b \Phi\right)+2 m^2 g_{a b} \Phi \Phi^{\dagger} ,
\end{split}
\end{equation}
where $\nabla_{(a}\xi_{b)}$ is the DeTurck term, $\xi^{a}=[\Gamma_{cd}^{a}(g)-\Gamma_{cd}^{a}(\bar{g})]g^{cd}$, and $\Gamma_{cd}^{a}(\bar{g})$ is the Levi-Civitta connection with respect of the reference metric $\bar{g}$.

At the point when $\xi^{a}=0$ , solutions derived from the Einstein-DeTurck equation correspond to those of our original equation. When $\xi^a \neq 0$, we encounter a Ricci soliton. However, it has been proved that these unwanted Ricci solitons do not exist in many circumstances if we have selected an appropriate reference metric \cite{Figueras:2011va,Figueras:2016nmo}. The condition $\xi^{a}=0$ actually implies a rewriting of the original Einstein equation (\ref{eq:Einsteineq}) in generalized harmonic coordinates $\Delta x^a=\Gamma_{c d}^a(\bar{g}) g^{c d}$. Hence, the choice of our reference metric $\bar{g}$ dictates the gauge choice. In order for DeTurck method to work, it is necessary for our reference metric $\bar{g}$ to possess the same asymptotic structure and gauge choice as the solution we aim to obtain. In the next subsection, we will determine the form of the reference metric.

The Einstein-DeTurck equation constitutes a well-defined elliptic problem which is more simplified for static solutions. In this work, we employ a Newton-Raphson method for solving, utilizing a fine non-uniform grid of 420×300 across the calculation domain [0,1]×[0,1] (represents the range of values for the coordinates $x$ and $y$, which will be introduced in the next subsection), ensuring precise numerical computation in critical areas while efficiently using computational resources. Results were cross-verified with closely matching outcomes on 400×400 and 450×450 grids (within error margins considered equivalent). Moreover, for the validity of results, we mandated that the relative error for static solutions be less than $10^{-5}$.

\subsection{Ansatzs and boundary conditions}
  We will focus on static axisymmetric solutions, and due to the DeTurck method, we need to give a most general ansatz under this spacetime symmetry. Thus, we first adjust our coordinate system to make $\partial_t$ and $\partial_\phi$ Killing fields. We will adopt the coordinate system constructed in \cite{Horowitz:2014gva}. Starting from the Poincaré coordinates $(t, r, z, \phi)$ of the pure AdS spacetime, the metric is 
\begin{equation}
\label{pure ads}
\mathrm{d} s^2=\frac{L^2}{z^2}\left[-\mathrm{d} t^2+\mathrm{d} r^2+r^2 \mathrm{~d} \phi^2+\mathrm{d} z^2\right].
\end{equation}
For numerical computations, the coordinate domain needs to be compactified since both $r$ and $z$ coordinates are $[0, \infty)$. We choose the following coordinate transformation
\begin{subequations}
\begin{align}
z=\frac{y \sqrt{2-y^2}}{1-y^2}\left(1-x^2\right) , 
\end{align}
\begin{align}
\quad r=\frac{y \sqrt{2-y^2}}{1-y^2} x \sqrt{2-x^2} ,
\end{align}
\end{subequations}
where both coordinates $x$ and $y$ are in range $[0, 1]$. We can also rewrite (\ref{pure ads}) in polar-like coordinates ($x$, $y$) as
\begin{equation}
\begin{split}
\mathrm{d} s^2=\frac{L^2}{\left(1-x^2\right)^2}&\left[-\frac{\left(1-y^2\right)^2\mathrm{d} t^2}{y^2\left(2-y^2\right)}+\frac{4\mathrm{d} x^2}{2-x^2}\right.\\
&\left.\quad+\frac{4\mathrm{d} y^2}{y^2\left(1-y^2\right)^2\left(2-y^2\right)^2}+x^2\left(2-x^2\right) \mathrm{d} \phi^2\right].
\end{split}
\end{equation}
Now we can give the most general metric ansatz for static and axisymmetric solutions
\begin{equation}
\begin{aligned}
\label{metric}
\mathrm{d} s^2=\frac{L^2}{\left(1-x^2\right)^2}\left[-\frac{\left(1-y^2\right)^2 Q_1 \mathrm{~d} t^2}{y^2\left(2-y^2\right)}+\frac{4 Q_4}{2-x^2}\left(\mathrm{~d} x+\frac{Q_3}{1-y^2} \mathrm{~d} y\right)^2\right. \\
\left.+\frac{4 Q_2 \mathrm{~d} y^2}{y^2\left(1-y^2\right)^2\left(2-y^2\right)^2}+x^2\left(2-x^2\right) Q_5 \mathrm{~d} \phi^2\right] ,\\
\end{aligned}
\end{equation}
where the $Q_i$, with $i \in\{1, \ldots, 7\}$ are the functions of $x$ and $y$ that we are tasked to determine through solving. We choose the reference metric with $Q_1=Q_2=Q_4=Q_5=1, Q_3=0$. For the maxwell field and the scalar field we take
\begin{equation}
\label{matter}
A=L Q_6\mathrm{~d}t, \quad \text { and } \Phi=\left(1-x^2\right)^{\Delta} y^{\Delta}\left(2-y^2\right)^{\frac{\Delta}{2}} Q_7,
\end{equation}
where $\triangle \equiv 3 / 2+\sqrt{9 / 4+L^2m^2}$. In \cite{Crisford:2017gsb}, similar results are obtained for different $\triangle$ values. To simplify our computational process, we only take into account the case where $\triangle=2$ in our work.

Subsequently, we establish the boundary conditions for the equation. For the comformal boundary located at $x=1$, the spacetime should be pure AdS, and we can obtain that
\begin{equation}
Q_1=Q_2=Q_4=Q_5=1, \quad Q_3=0, \quad \text { and } \quad Q_6=a\left(1-y^2\right)^n,
\end{equation}
where $a$ is an amplitude, $n$ determines the decay behavior of the potential at the boundary. In our work, similar results are obtained for different $n$ values. Along the symmetric axis located at $x=0$, we impose
\begin{equation}
\frac{\partial Q_1}{\partial x}=\frac{\partial Q_2}{\partial x}=\frac{\partial Q_4}{\partial x}=\frac{\partial Q_5}{\partial x}=\frac{\partial Q_6}{\partial x}=0, \quad Q_4=Q_5, \quad \text { and } \quad Q_3=0.
\end{equation}
At $y=0$, which means $r=z=0$, corresponding to the intersection point of the conformal boundary and the axis of symmetry, we find
\begin{equation}
Q_1=Q_2=Q_4=Q_5=1, \quad Q_3=0, \quad \text { and } \quad Q_6=0.
\end{equation}
At the $y=0$ where the Poincaré horizon locates, the boundary conditions are
\begin{equation}
Q_1=Q_2=Q_4=Q_5=1, \quad Q_3=0, \quad \text { and } \quad Q_6=0,
\end{equation}

After specifying the boundary conditions and assuming the ansatz, we can now solve the equation of motion (\ref{eqs:EOM}). For the equation (\ref{eq:KGseq}), there are multiple solutions. While \cite{Crisford:2017gsb} explored the ground state solutions, we will proceed to solve for the case of first and second excited states, and investigate their properties.

\section{The results}\label{sec3}
  In this section, we firstly consider the scalar field as a test field, finding excited state solutions on a fixed spacetime background and noting differences from the ground state. Subsequently, we couple the scalar field to the Einstein-Maxwell field and solve the coupled system of nonlinear partial differential equations (\ref{eqs:EOM}) with the ansatzs. We investigate whether a lower bound on charge that protects the WCCC still exists for excited-state scalar fields.

\subsection{Linear Solutions}
  When the scalar field is treated as a test field, terms related to the scalar field in the Einstein equations vanish, (\ref{eqs:EOM}) simplifying to the Einstein-Maxwell-$\Lambda$ equation. In such a fixed spacetime background, considering only static solutions, the equation (\ref{eq:KGseq}) reduces to a linear equation

\begin{equation}
\label{linear eq}
\left(\nabla_a \nabla^a-m^2\right) \Phi=q^2 A_a A^a \Phi .
\end{equation}
Since we have already chosen the value for $\triangle$, the value of $m$ is also determined. So this equation can essentially be considered an eigenvalue equation where $q^2$ is the eigenvalue and $\Phi$ is the eigenfunction. In \cite{Crisford:2017gsb} they solved for the smallest eigenvalue $q_{min}$ of the equation and demonstrated that a scalar field with a charge value greater than $q_{min}$ would induce instability in the Einstein-Maxwell background. This instability implies the formation of a non-zero scalar field, which necessitates the full resolution of the coupled equations of motion (\ref{eqs:EOM}). 

\begin{figure}[H]
\centering

\begin{subfigure}{0.48\textwidth}
  \centering
  \includegraphics[width=\linewidth]{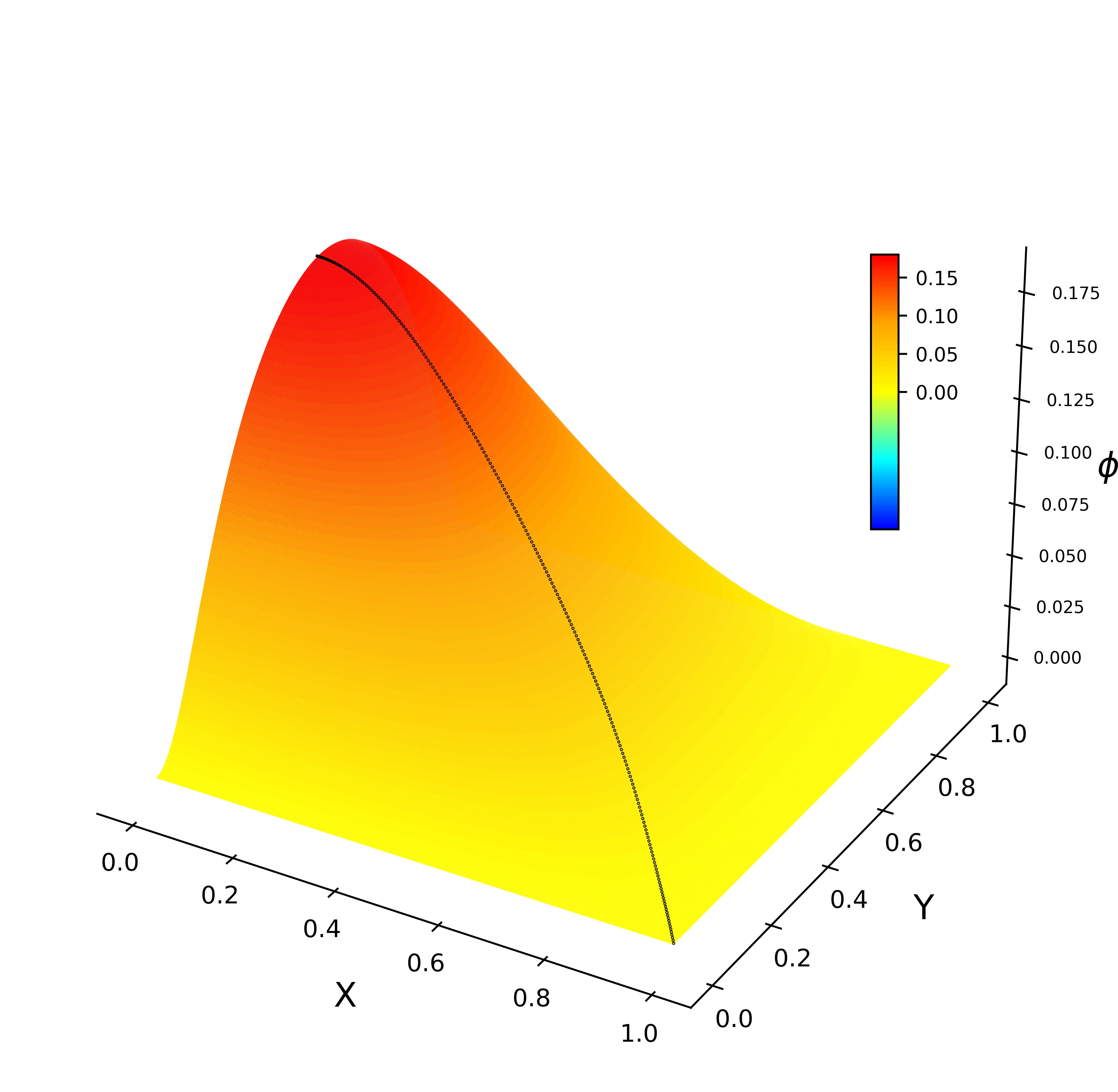}
  \caption{}
  \label{fig:linearsub1}
\end{subfigure}\hfill
\begin{subfigure}{0.48\textwidth}
  \centering
  \includegraphics[width=\linewidth]{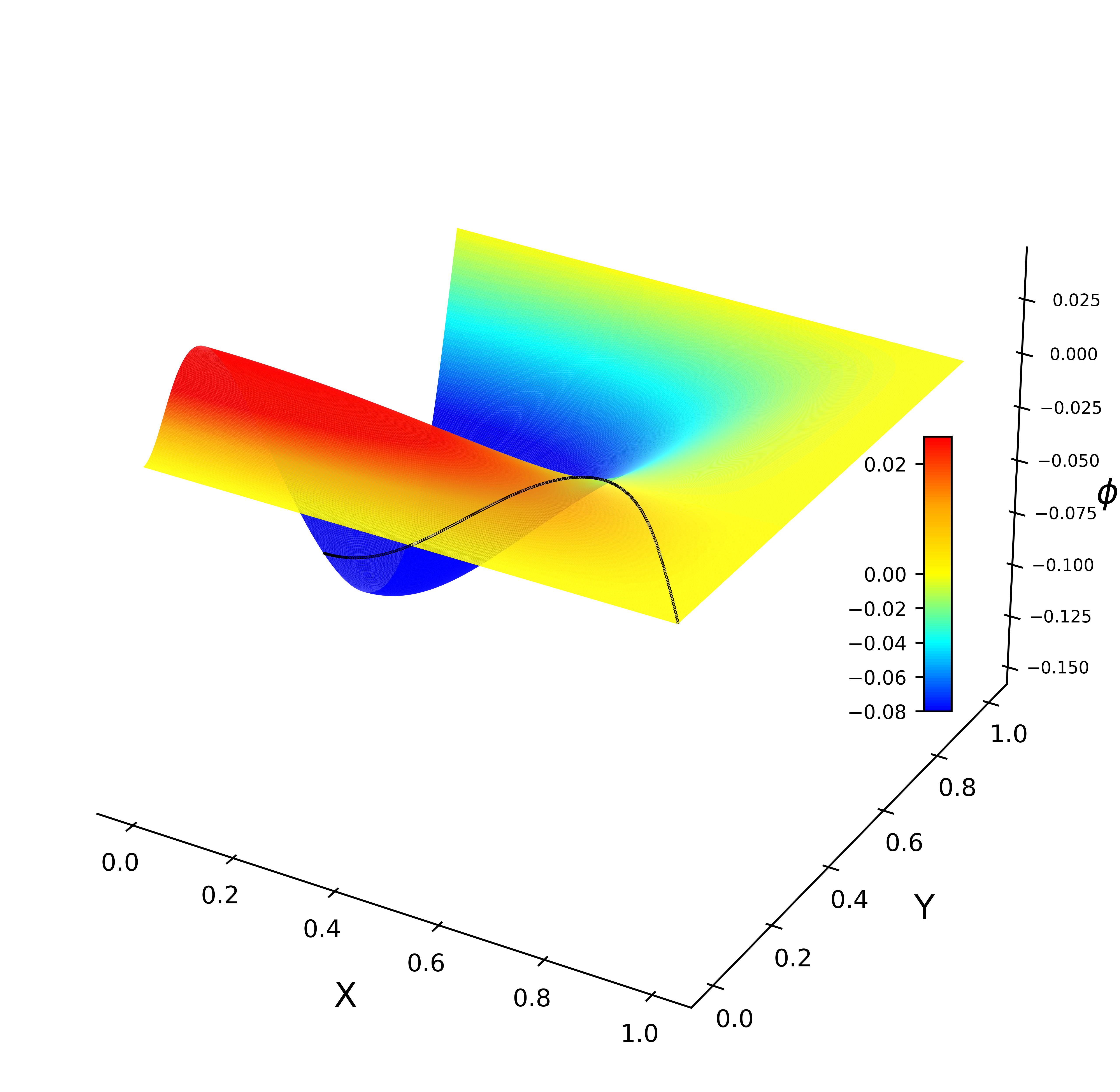}
  \caption{}
  \label{fig:linearsub2}
\end{subfigure}\hfill
\begin{subfigure}{.5\textwidth}
  \centering
  \includegraphics[width=\linewidth]{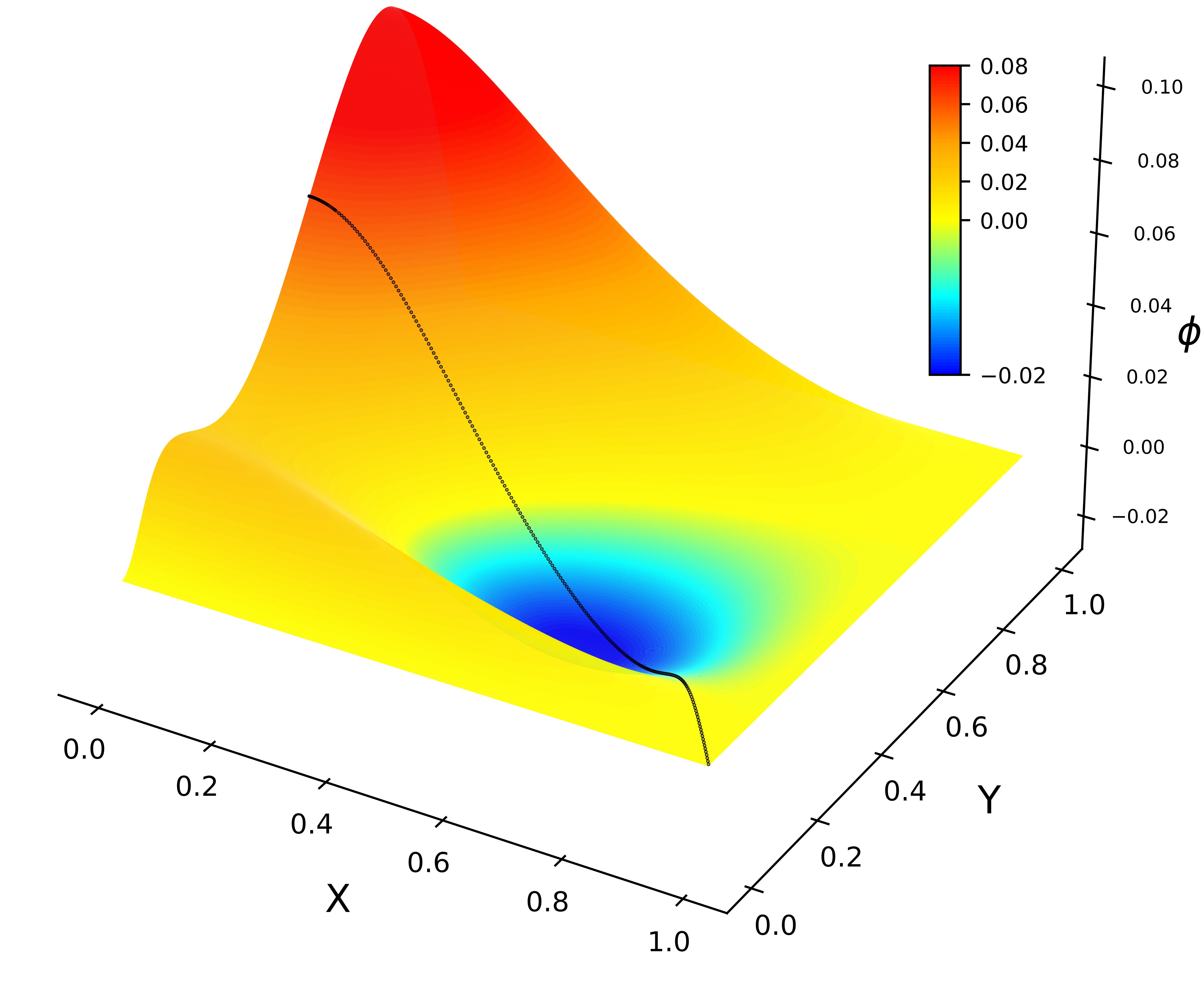}
  \caption{}
  \label{fig:linearsub3}
\end{subfigure}
\caption{Distribution of scalar fields as a function of the coordinates x and y when $a = 6$, $n = 8$, with figures \textbf{(a)}, \textbf{(b)}, and \textbf{(c)} representing the ground state, first excited state, and second excited state of the scalar field, respectively. The black lines on all three surfaces denote the curve $\ell$.}
\label{fig:linearphi}
\end{figure}

Before solving these nonlinear partial differential equations, let us revisit the equation (\ref{linear eq}). This equation should have multiple distinct eigenvalues $q$ and eigenfunctions $\Phi$. We proceed to solve for the second and third smallest eigenvalues ($q_{2nd}, q_{3rd}$), corresponding to the first and second excited states of the scalar field solutions. We present the distribution of scalar fields corresponding to $q_{min}$, 
 $q_{2nd}$, and $q_{3rd}$, as shown in Fig.\ref{fig:linearphi}. The black lines on the distributions mark the curve $\ell$ : $f_{\ell} (x,y)=0$ \footnote{We choose $f_{\ell}(x,y)=\frac{y \sqrt{2-y^2}}{1-y^2}-(1-x^2)$.}. It can be seen that these solutions exhibit completely different configurations. For the smallest eigenvalue (Fig.\ref{fig:linearsub1}), the value of $\Phi$ is positive across the coordinate domain, with no intersection between the curve $\ell$ and the $\Phi=0$ plane, which we refer to as the ground state solution of the scalar field.
 
For the solution with the second smallest eigenvalue (Fig.\ref{fig:linearsub2}), the value of $\Phi$ at $y=0$ approaches a positive zero, and at $y=1$ it approaches a negative zero. There is only one point of intersection between the curve $\ell$ and the $\Phi=0$ plane, which corresponds to the first excited state of the scalar field. For the solution with the third smallest eigenvalue (Fig.\ref{fig:linearsub3}), the value of $\Phi$ is less than zero within a closed region and greater than zero outside this region. It is clearly observed that the curve $\ell$ intersects the $\Phi=0$ plane twice, corresponding to the second excited state of the scalar field.

\begin{figure}[H]
\centering
\includegraphics[width=.7\textwidth]{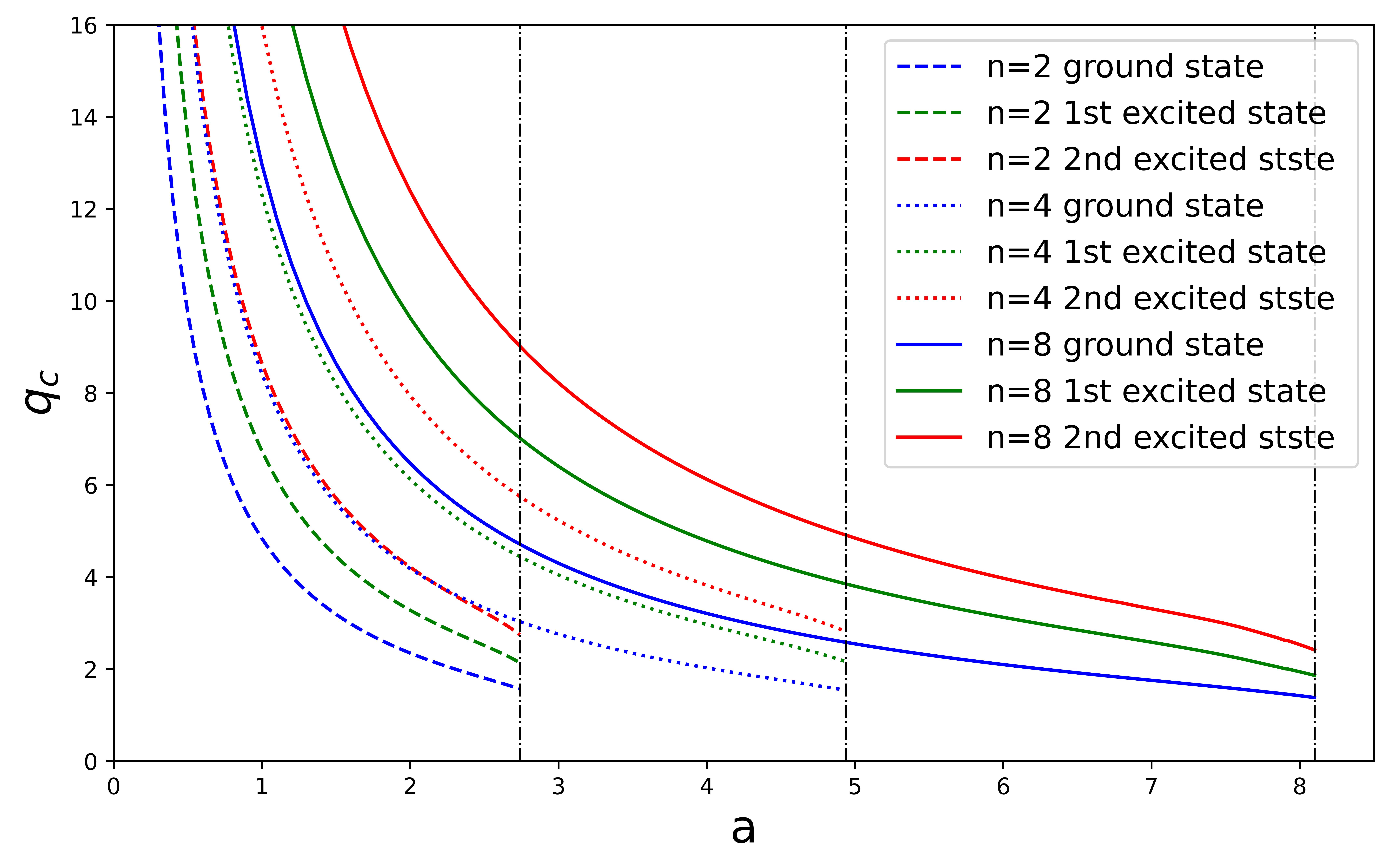}\hfill
\caption{The critical charge $q_c$ as a function of the amplitude $a$ is shown for the ground state \textbf{(blue)}, first excited state \textbf{(green)}, and second excited state \textbf{(red)} scalar fields. The \textbf{dashed}, \textbf{dotted}, and \textbf{solid} lines represent the cases for $n = 2$, $n = 4$ and $n = 8$  respectively. The black dash-dotted line indicates the value of $a = a_{max}$ (2.74, 4.95, 8.08) at which a naked singularity appears.}
\label{fig:qc}
\end{figure}
It has been proven that $q_{min}$ signifies the critical charge for the formation of a non-zero scalar field in \cite{Crisford:2017gsb}. We believe that $q_{2nd}$ and $q_{3rd}$ also represent the critical charges corresponding to the formation of excited state scalar fields. We will see this in the next subsection. Therefore, we now refer to $q_{min}$, $q_{2nd}$ and $q_{3rd}$ collectively as critical charges $q_c$. As the amplitude $a$ varies, the fixed background spacetime also changes. Consequently, for each distinct value of $a$, we can calculate the corresponding critical charge $q_c$. In Fig.\ref{fig:qc}, we show how the values of $q_c$ for the ground state (blue), first (green) and second excited states (red) vary with the amplitude $a$ for different $n$. The excited state solutions exhibit the same trend as the ground state, where $q_c$ is a decreasing function of amplitude $a$. This is anticipated since an increase in $a$ effectively represents an increase in the electric field, allowing the excited scalar field to carry a smaller charge to form. From the graph, it can be observed that for $n = 2, 4, 8$, both the ground state solutions and excited state solutions correspond to the same maximum value of $a$ (respectively equal to 2.74, 4.95, and 8.08), denoted as $a_{max}$ (marked by black dash dot line). This is understandable since the spacetime background is fixed, $a_{max}$ corresponds to the value of $a$ when the naked singularity appears in the Einstein-Maxwell background.

\begin{figure}[htbp]
\centering
\includegraphics[width=.5\textwidth]{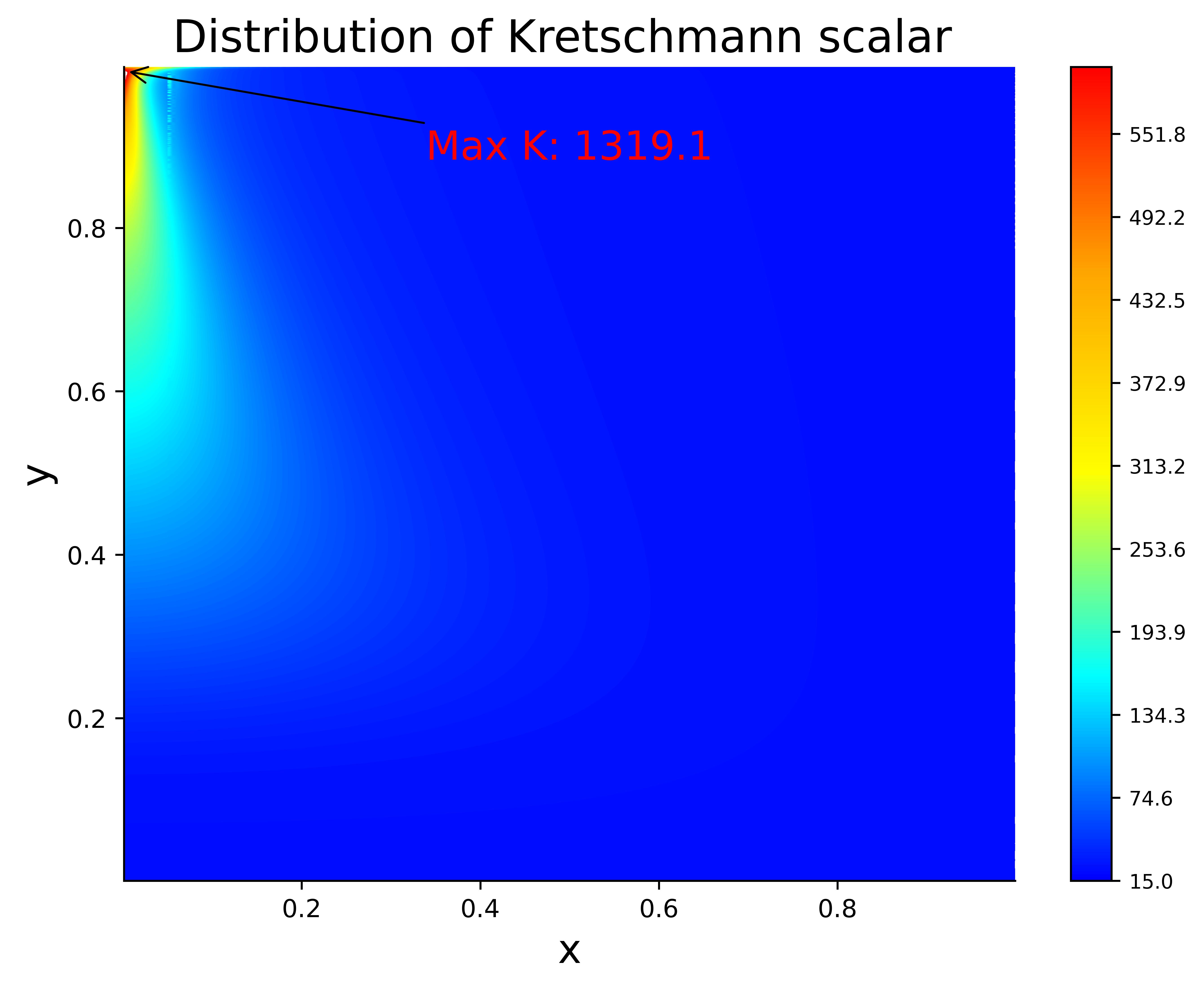}\hfill
\includegraphics[width=.5\textwidth]{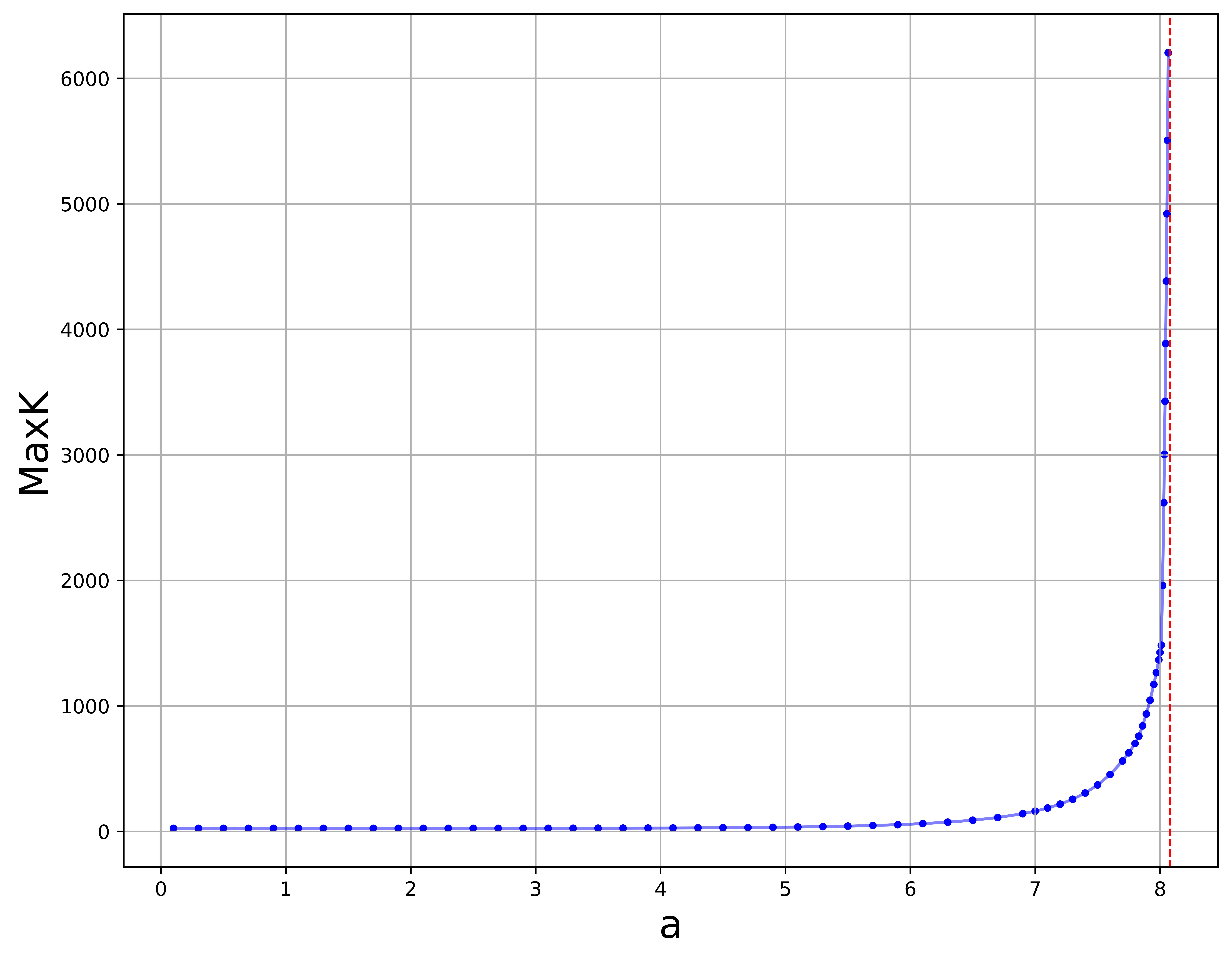}
\caption{\textbf{Left}: Isometric distribution map of the Kretschmann scalar when $a = 8$, $n= 8$. \textbf{Right}: Maximum value of Kretschmann scalar over the Coordinate domain as a function of $a$ with $n = 8$, where red dashed line represents $a = a_{max} = 8.08$.}
\label{fig:K}
\end{figure}

The reason why we think the naked singularity appears at $a = a_{max}$ is displayed in Fig.\ref{fig:K}. In the left panel of Fig.\ref{fig:K}, we display an isometric distribution map of the Kretschmann scalar $K \equiv L^4 R_{a b c d} R^{a b c d}$ in the bulk when $ a= 8$, $n = 8$. Since  $ a= 8$ is quite close to $a_{max} = 8.08$, it is clearly observable that there is an area outside the Poincaré horizon with significant curvature. In the right panel of Fig.\ref{fig:K}, we present the variation of the maximum value of the Kretschmann scalar in the bulk as a function of $a$. It is easily noticeable that the Kretschmann scalar appears to blow up when amplitude $a$ approach a certain critical value $a_{max} = 8.08$ (marked by red dashed line). Hence, we infer the emergence of a naked singularity when $a = a_{max}$. It is worth mentioning that our obtained $a_{max}$ is slightly larger than the result in \cite{Crisford:2017gsb}, possibly due to differences in computational accuracy caused by grid precision. In the next subsection, we will demonstrate that the presence of a non-zero scalar field means that $a$ exceeding $a_{max}$ does not necessarily indicate the emergence of a naked singularity. Naked singularities will not appear when the scalar field's charge is greater than a lower bound $q_c^{bound}$, and this conclusion remains valid for excited state scalar fields. This suggests the existence of a lower bound for the charge carried by the scalar field to prevent the violation of the Weak Cosmic Censorship Conjecture (WCCC).

\subsection{Nolinear Solutions }

  In this subsection, we will fully solve the equation of motion (\ref{eqs:EOM}) and obtain hairy solutions for the excited state scalar field. In fact, due to the AdS/CFT correspondence \cite{Maldacena:1997re,Gubser:1998bc,Witten:1998qj,Aharony:1999ti}, the model in AdS spacetime that couple gravity with a Maxwell field and a scalar field can be used to construct the simplest holographic superconductor model \cite{Hartnoll:2008vx,Hartnoll:2008kx} (known as holographic s-wave superconductor models \cite{Cai:2015cya}). In AdS/CFT correspondence, the scalar field in the bulk gravitational theory is dual to the gauge-invariant operator in the  Quantum Field Theory (QFT) on boundary. For the form of the scalar field $\Phi$ we have chosen, the expectation value of its dual operator is $\left\langle\mathcal{O}_2\right\rangle=\left(1-y^2\right)^2 Q_7(y, 1)$. In the following analysis, $\left\langle\mathcal{O}_2\right\rangle$ will help us better understand the distribution of the scalar field and the impact of changes in the amplitude $a$ on the scalar field.

\begin{figure}[H]
\centering

\includegraphics[width=.5\textwidth]{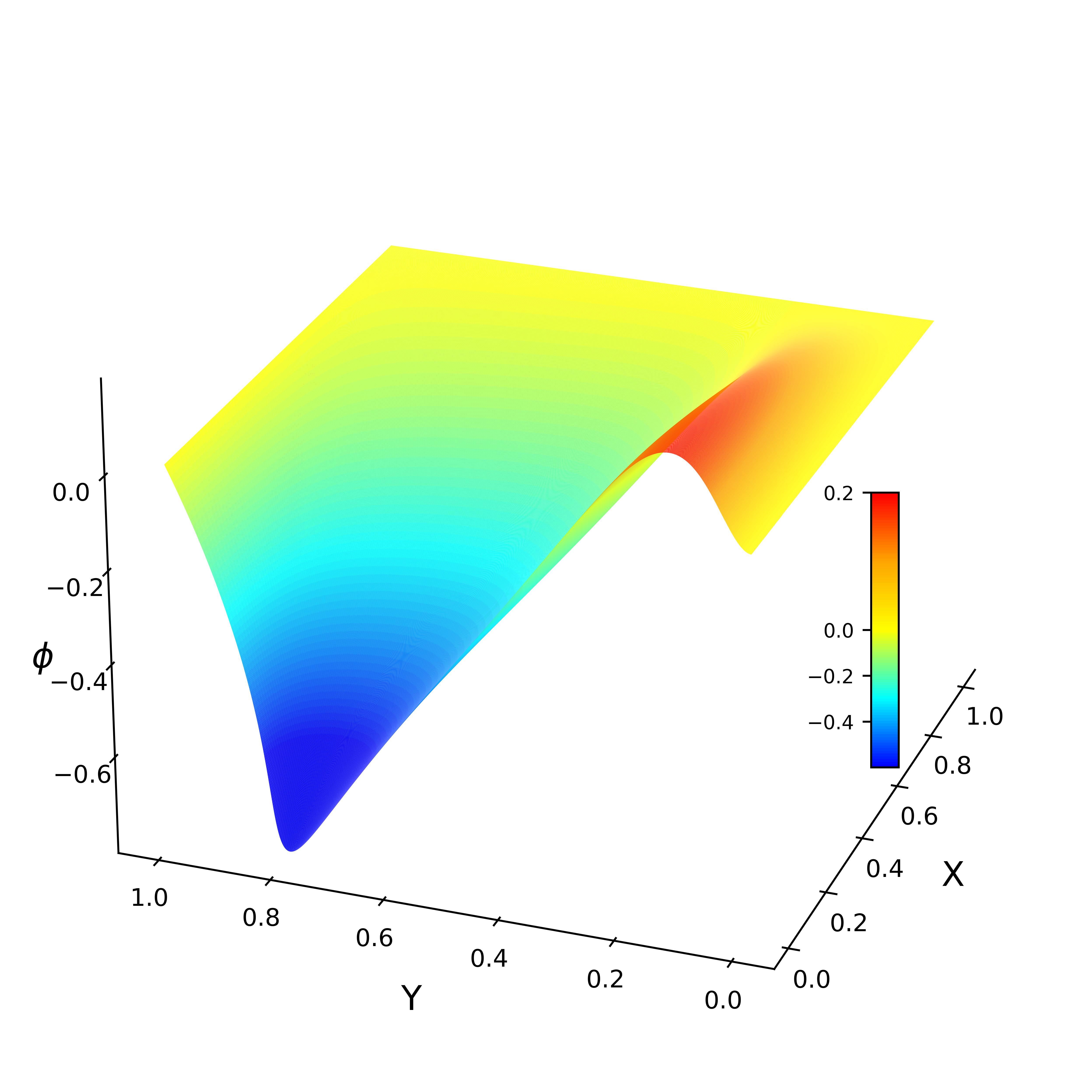}\hfill
\includegraphics[width=.5\textwidth]{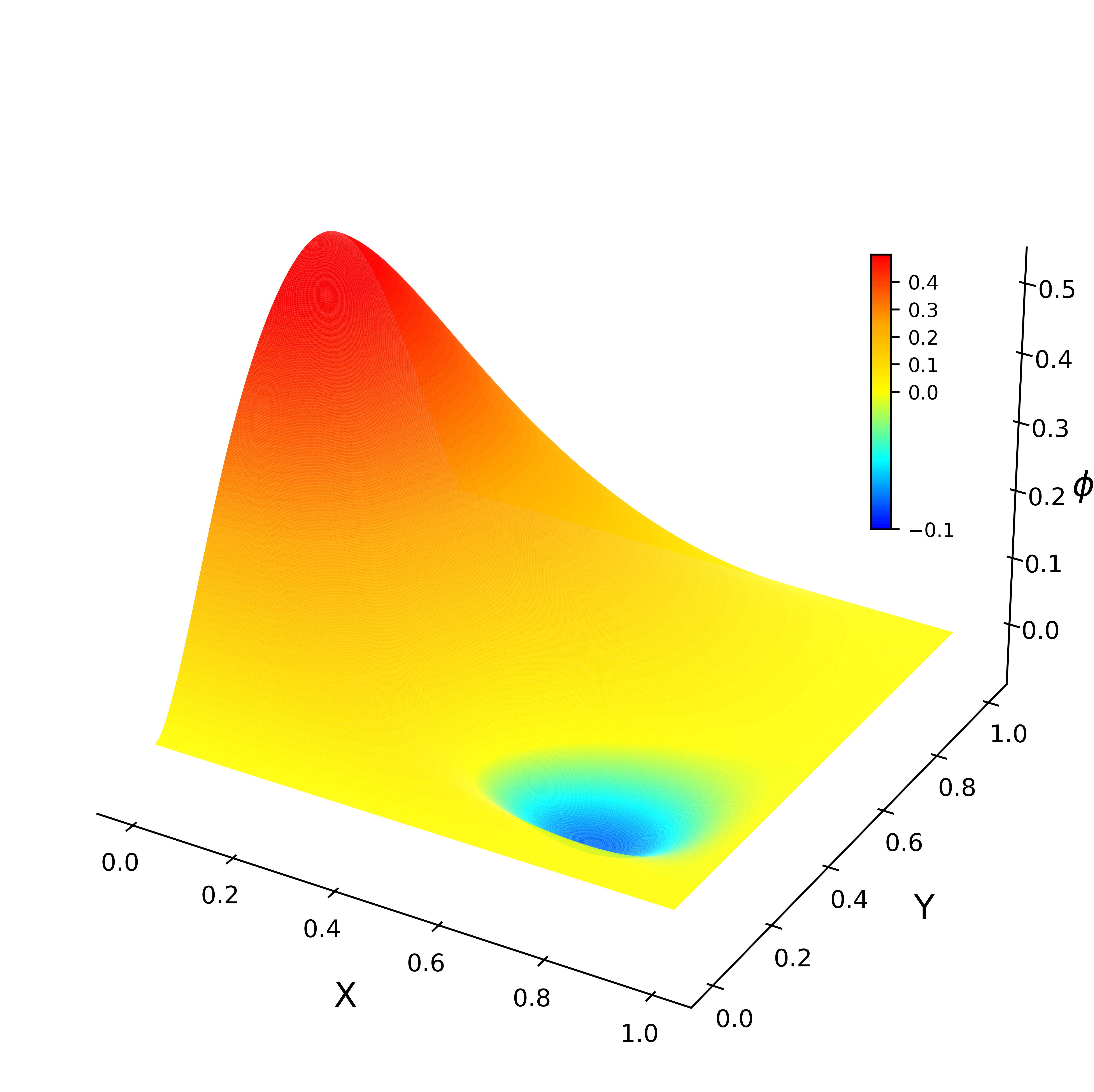}
\caption{\textbf{Left}: Distribution of scalar fields when $a = 8.5$, $q = 2.9$ for the first excited-state solution. \textbf{Right}:  Distribution of scalar fields when $a = 7.7$, $q = 4$ for the second excited-state solution.}
\label{fig:nolinear}
\end{figure}

We now display the distribution of the excited state scalar fields obtained from fully solving (\ref{eqs:EOM}) in Fig.\ref{fig:nolinear}. The left panel presents the distribution of the first excited state scalar field when $a = 8.5$ and $q = 2.9$ (with a different viewing angle than Fig.\ref{fig:linearsub2}). The right panel shows the distribution of the second excited state scalar field at $a = 7.7$ , $q = 4$. It is observed that the results we obtained have a similar configuration to those from the linear solutions in the previous subsection, indicating that the solutions we have found indeed correspond to the anticipated excited-state scalar field solutions.

\begin{figure}[htbp]
\centering

\includegraphics[width=.5\textwidth]{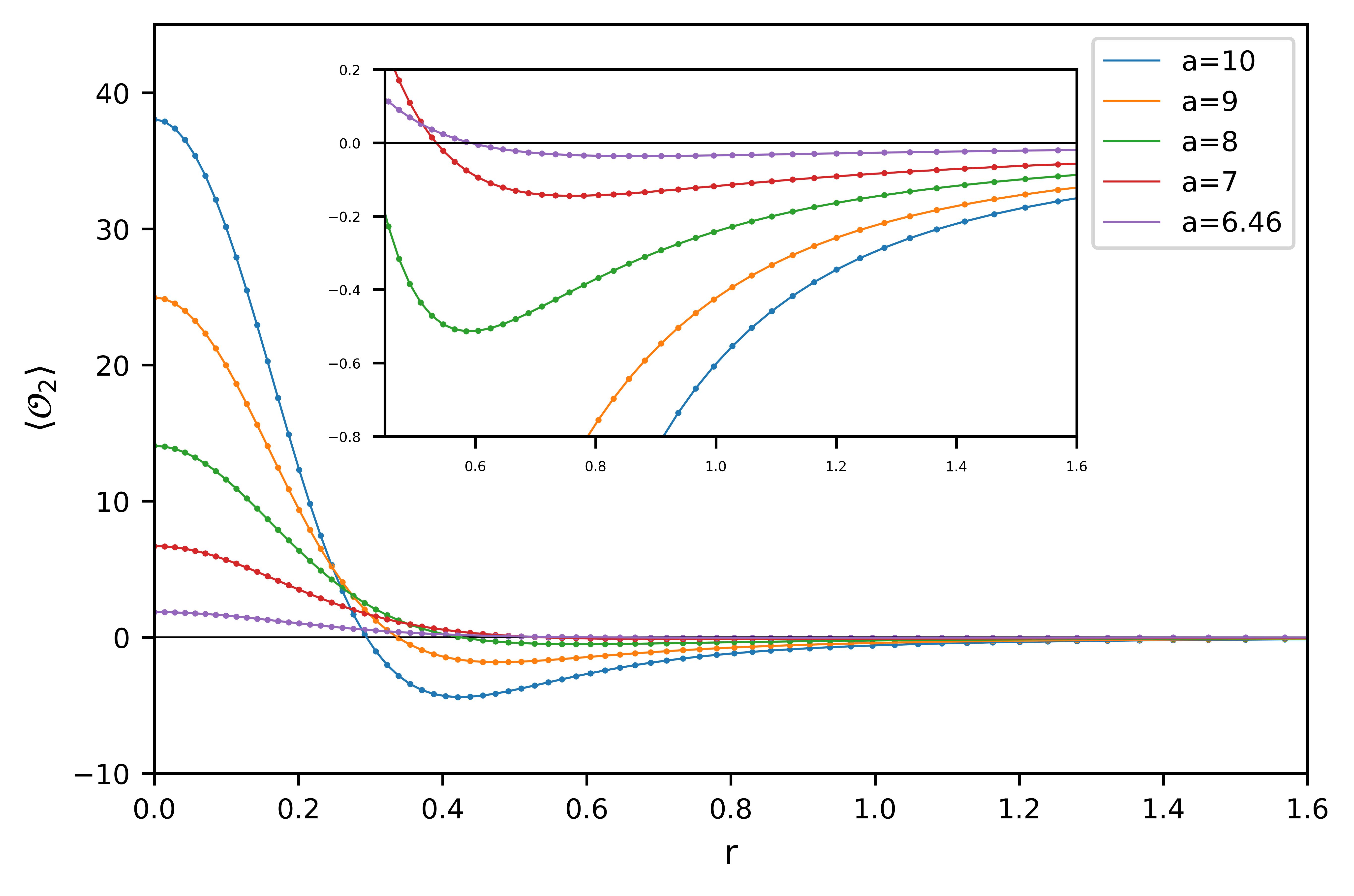}\hfill
\includegraphics[width=.4\textwidth]{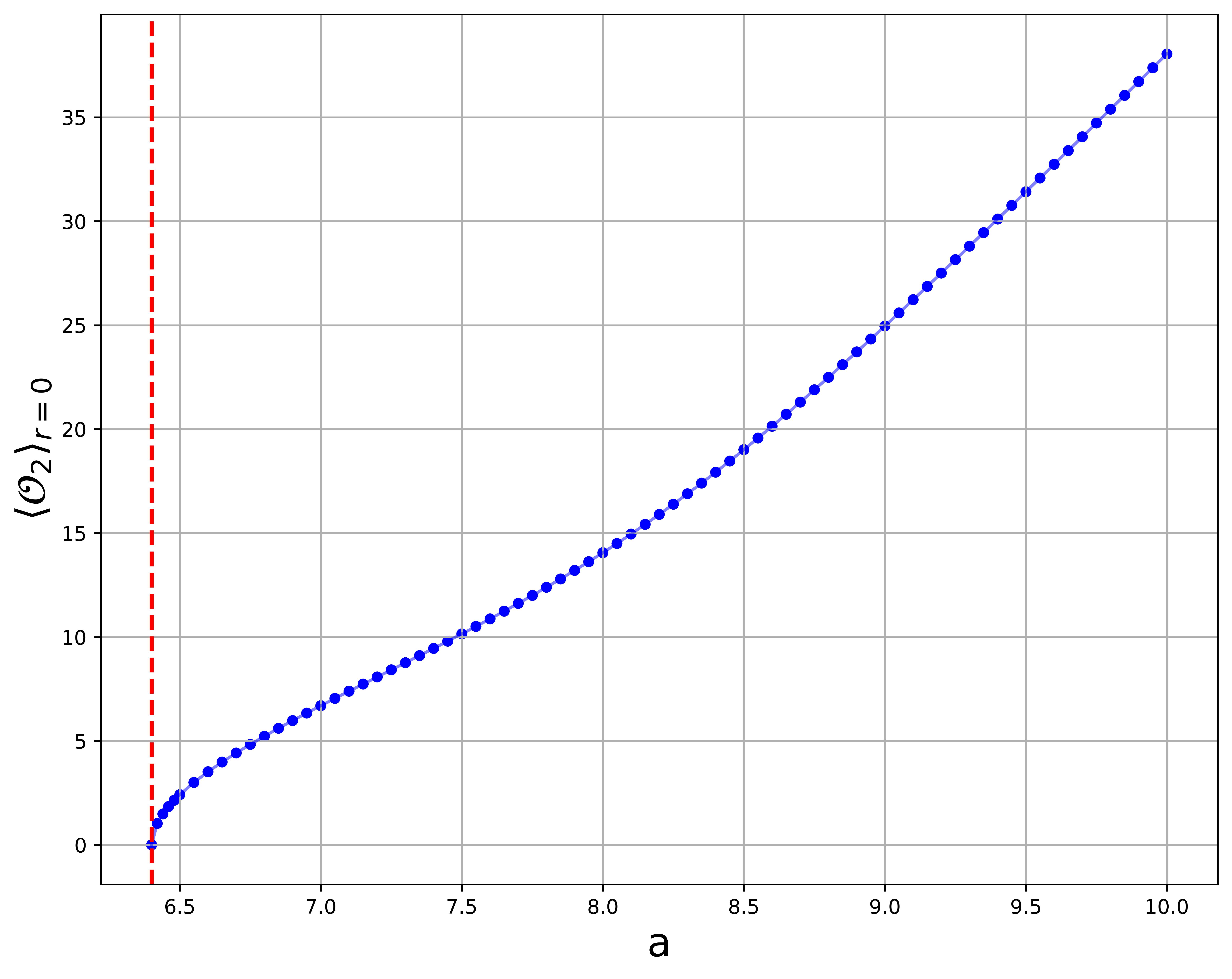}
\caption{\textbf{Left}: For the first excited scalar field  $\langle\mathcal{O}_2\rangle$ as $a$ function of $r$, for several values of a and with $n = 8$, $q = 2.9$ ; from top to bottom we have $a =$ 10, 9.0, 8.0, 7.0, 6.46. \textbf{Right}: $\langle\mathcal{O}_2\rangle$ at $r = 0$ plotted as a function of amplitude $a$ for $q = 2.9$, with the vertical dashed red line representing $a = 6.4$, which corresponds to the value of $a$ at $q_c = 2.9$ in the $q_c - a$ curve for the first excited scalar field calculated in the previous subsection.}
\label{fig:O2 first}
\end{figure}

Now let's take a look at how the excited-state scalar condensate evolves with the growth of $a$ at this moment. In the left panel of Fig.\ref{fig:O2 first}, we depict the functional relationship between the expectation value of the dual operator $\left\langle\mathcal{O}_2\right\rangle$ of the first excited-state scalar field and the boundary radial coordinate $r$ for several $a$ values, at $n = 8$ and $q = 2.9$. From the panel, the configuration of the first excited state is evident. All functions of $\left\langle\mathcal{O}_2\right\rangle$ intersect the $\left\langle\mathcal{O}_2\right\rangle = 0$ axis only once, with the maximum function value occurring at $r = 0$ and tending towards zero at infinity. The curves in the panel, from top to bottom, correspond to $a =$ 10, 9, 8, 7, 6.46 respectively. It can be observed that as $a$ increases, the maximum value of $\left\langle\mathcal{O}_2\right\rangle$ also increases.
To illustrate this point more clearly, in the right panel of Fig.\ref{fig:O2 first}, we depict the relationship between the value of $\left\langle\mathcal{O}_2\right\rangle$ at $r = 0$ and the variation in $a$ when $q = 2.9$. The vertical dashed line represents $a = 6.4$, which corresponds to the value of $a$ at $q_c = 2.9$ in the $q_c - a$ curve calculated in the previous subsection. From this panel, we can observe the scalar condensate begin to occur at $a = 6.4$. Moreover, with the continuous increase in $a$ (even beyond $a_{max}$), we still obtain smooth solutions without naked singularities in our calculation domain. Additionally, it seems that as $a$ becomes arbitrarily large, corresponding significant scalar hair condenses to maintain the nonsingularity of spacetime. The Weak Cosmic Censorship Conjecture (WCCC) appears to be potentially rescued due to the emergence of the scalar field. Another important point to note is that the scalar condensation 
 begin to occur precisely at the value of $a$ corresponding to the critical charge $q_c = 2.9$ of the first excited state scalar field. This indicates that the $q_c - a$ curve obtained in the preceding subsection indeed represents the critical line for the appearance of the first excited-state scalar field.

\begin{figure}[htbp]
\centering

\includegraphics[width=.5\textwidth]{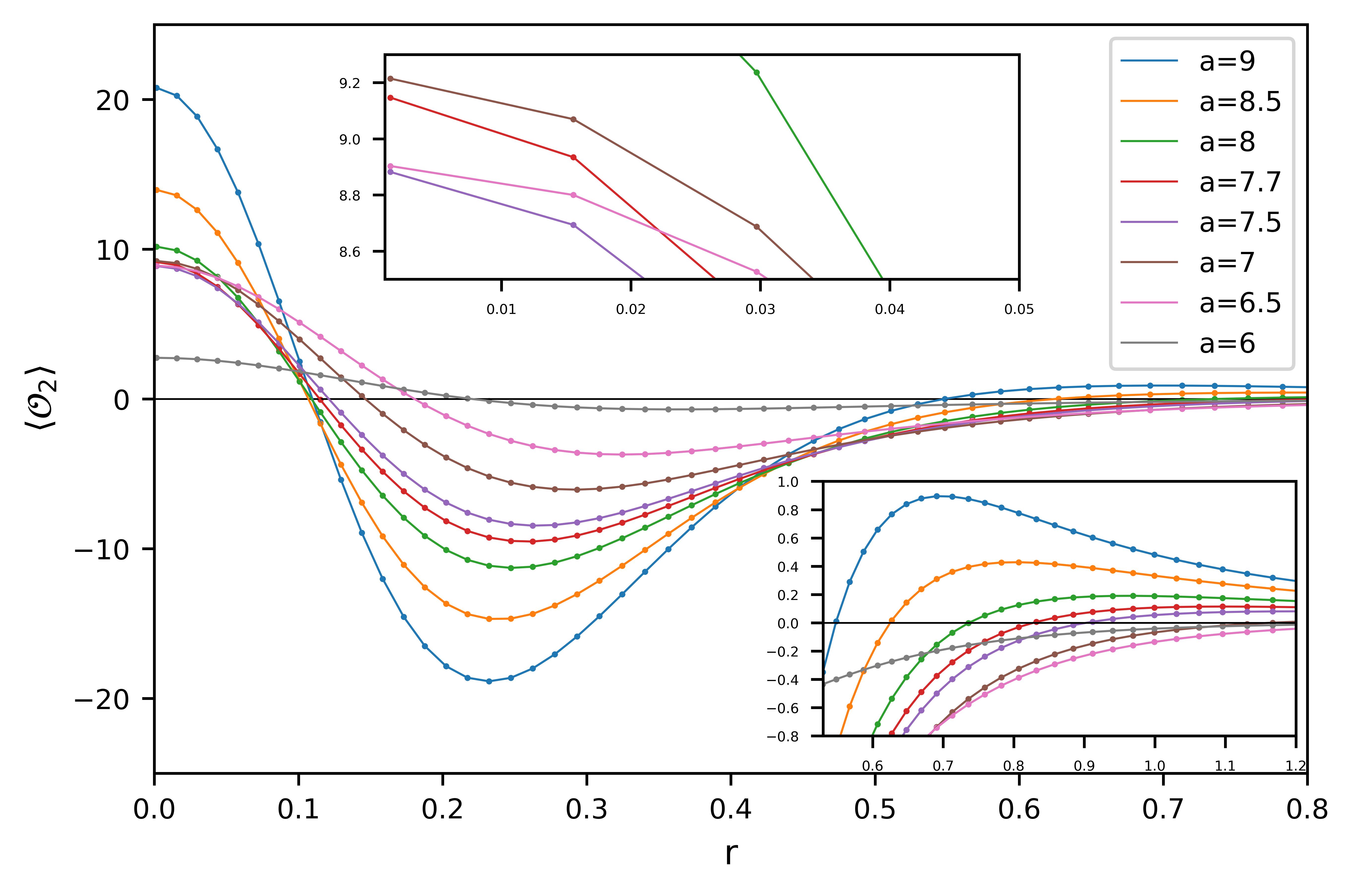}\hfill
\includegraphics[width=.4\textwidth]{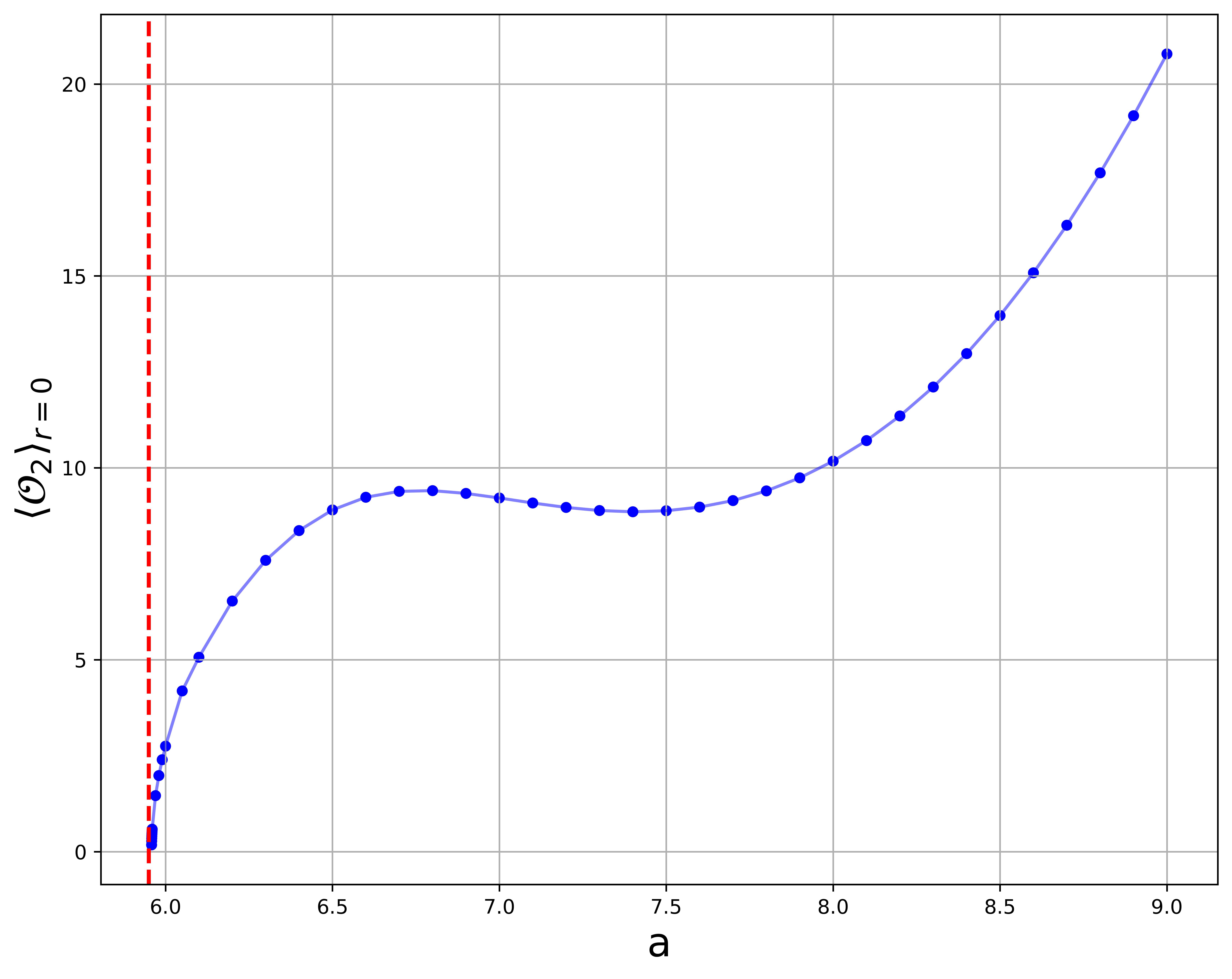}
\caption{\textbf{Left}: For the second excited scalar field $\langle\mathcal{O}_2\rangle$ as $a$ function of $r$, for $a =$ 9.0, 8.5, 8.0, 7.7, 7.5, 7.0, 6.5, 6.0 and with $n = 8$, $q = 4$. \textbf{Right}: $\langle\mathcal{O}_2\rangle$ at r = 0 plotted as a function of amplitude $a$ for $q = 4$, with the vertical dashed red line representing $a = 5.95$, which match the value of $a$ at $q_c = 4$ in the $q_c - a$ curve for the second excited scalar field.}
\label{fig:O2 second}
\end{figure}

Next, we present the scenario of the second excited-state scalar field. In the $\langle\mathcal{O}_2\rangle - r$ graph shown on the left panel of Fig.\ref{fig:O2 second}, at $n = 8$ and $q = 4$ for several amplitude $a =$ 9.0, 8.5, 8.0, 7.7, 7.5, 7.0, 6.5, 6.0. The graph distinctly displays the configuration of the second excited state, where the curves intersect the $\langle\mathcal{O}_2\rangle = 0$ axis twice and tend towards zero at infinity. However, unlike the ground state and the first excited state, there is also a significant distribution of scalar condensate away from $r = 0$. This leads to the non-monotonic increase of $\langle\mathcal{O}_2\rangle$ at $r = 0$ with respect to $a$, as depicted in the right panel of Fig.\ref{fig:O2 second}. Nonetheless, $\langle\mathcal{O}_2\rangle$ at $r = 0$ continues to represent the extent of scalar condensate to a certain degree. For any arbitrarily large $a$, we can still anticipate a correspondingly significant scalar condensate to maintain the Weak Cosmic Censorship Conjecture (WCCC). Similar to the first excited state, the appearance of scalar condensate for the second excited state at $a = 5.95$ precisely matches the $a$ value corresponding to the critical charge of the second excited state, $q_c = 4$. Therefore, the $q_c - a$ curve for the second excited state indeed corresponds to the critical curve for the appearance of the second excited-state scalar field.

We observe that scalar condensate increases correspondingly with the increase in $a$. Consequently, the scalar hair within the bulk increases as the spatial electric field grows, preventing the appearance of naked singularities. However, when we lower the scalar field charge $q$ to reduce its interaction strength with the electric field, can we still achieve the same results? Is there a critical value of charge at which, even with the presence of a scalar field, naked singularities cannot be prevented if the charge carried by the scalar field is less than that critical value? To explore this further, we will examine the solutions for the first and second excited states while keeping their $a$ values fixed and gradually reducing the excited-state scalar fields charges $q$. We will observe whether naked singularities occur in space (relevant studies have already been conducted for the ground state solution in \cite{Crisford:2017gsb}. It turns out there is indeed a critical charge. When the charges carried by the excited-state scalar fields is less than this critical value $q_c^{singular}$, spacetime curvature can become arbitrarily large.

\begin{figure}[htbp]
\centering

\includegraphics[width=.45\textwidth]{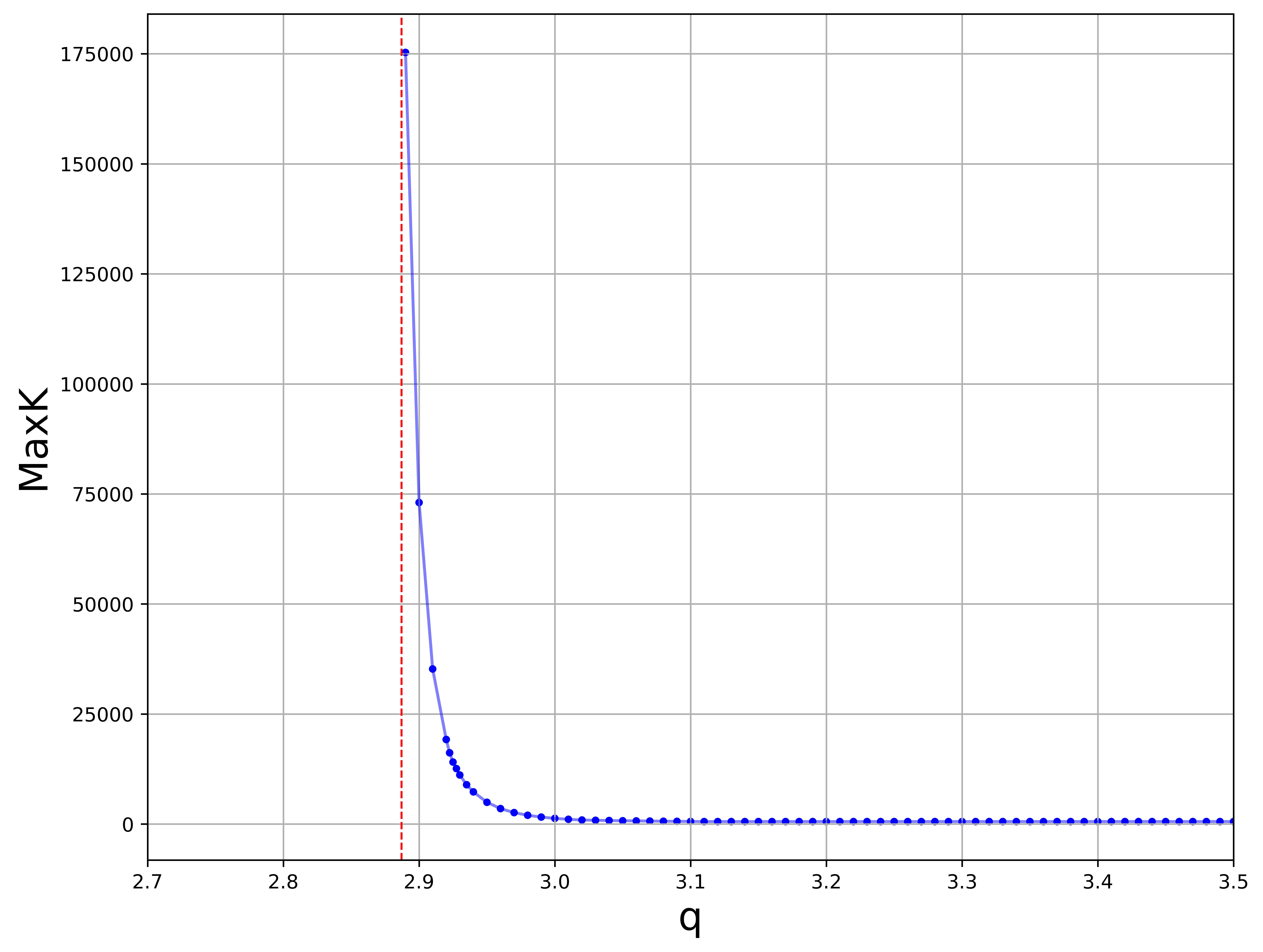}\hfill
\includegraphics[width=.45\textwidth]{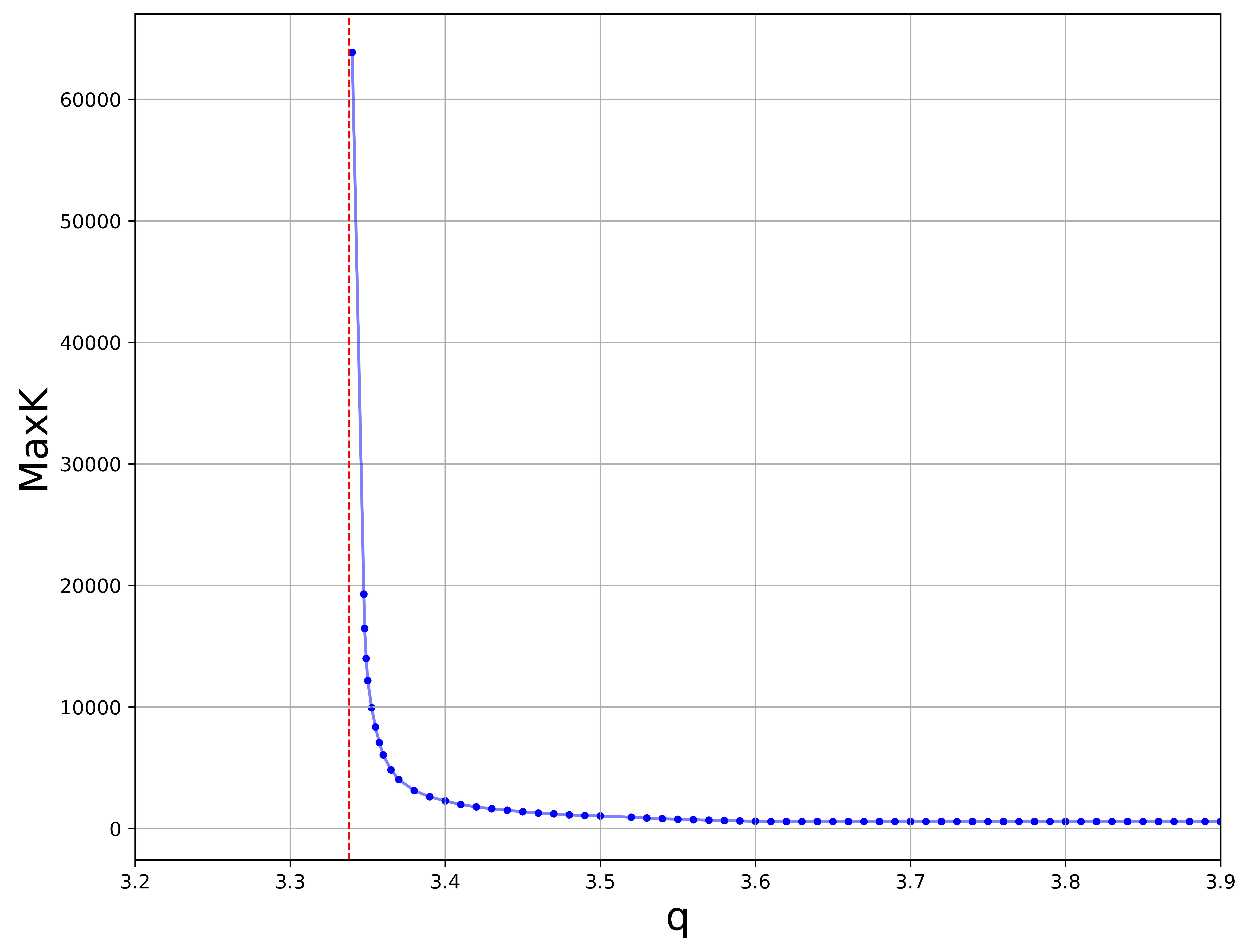}
\caption{\textbf{Left}: Maximum of Kretschmann scalar over our calculation domain for the first excited-state solution as a function of $q$ for fixed $a = 8.5 > a_{max}$ $n = 8$ and $\Delta = 2$, where red dashed line represents $q = q_c^{singular} = 2.88$. \textbf{Right}: The second excited-state solution case fixed $a = 7.7 < a_{max}$, where red dashed line represents $q = q_c^{singular} = 3.34$.}
\label{fig:nolinear maxk}
\end{figure}

In Fig.\ref{fig:nolinear maxk}, we present these results. The panels illustrate the maximum value of Kretschmann scalar over our integration domain concerning the change in charge $q$ for $n = 8$. The left panel represents the scenario with a fixed $a = 8.5$, corresponding to the first excited-state solution. The right panel represents the case with $a = 7.7$, corresponding to the second excited-state solution. It is clearly visible that when the charge value falls below a certain critical value $q_c^{singular}$, Kretschmann scalar diverges. In the panel, this is denoted by the vertical red dashed line ($q = 2.88$ in the left panel and $q = 3.34$ in the right panel). This signifies the appearance of naked singularities, indicating that despite the existence of excited-state scalar fields, when the charge is sufficiently low, the Weak Cosmic Censorship Conjecture (WCCC) is still violated.

\begin{figure}[htbp]
\centering

\includegraphics[width=.5\textwidth]{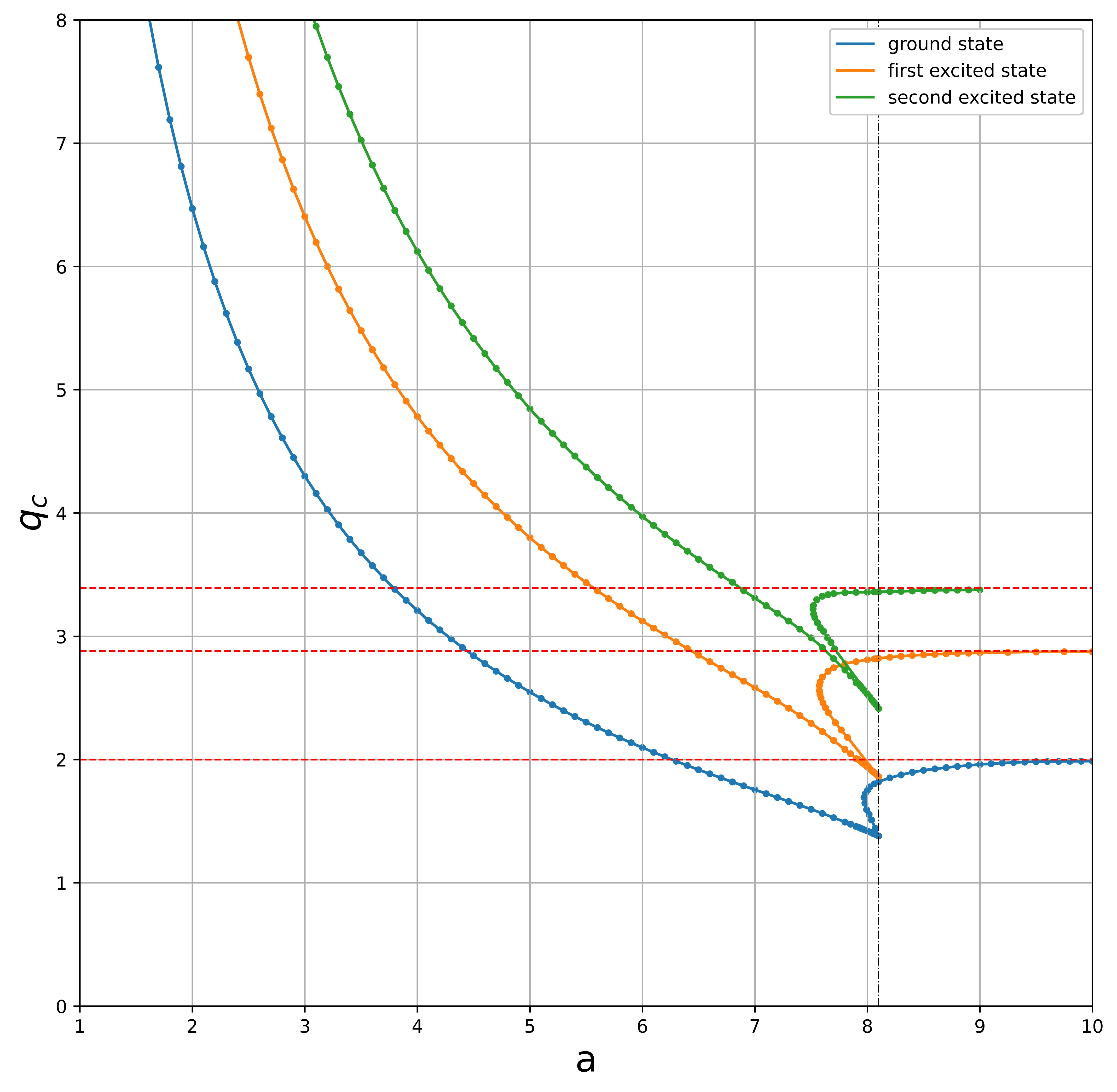}\hfill
\includegraphics[width=.5\textwidth]{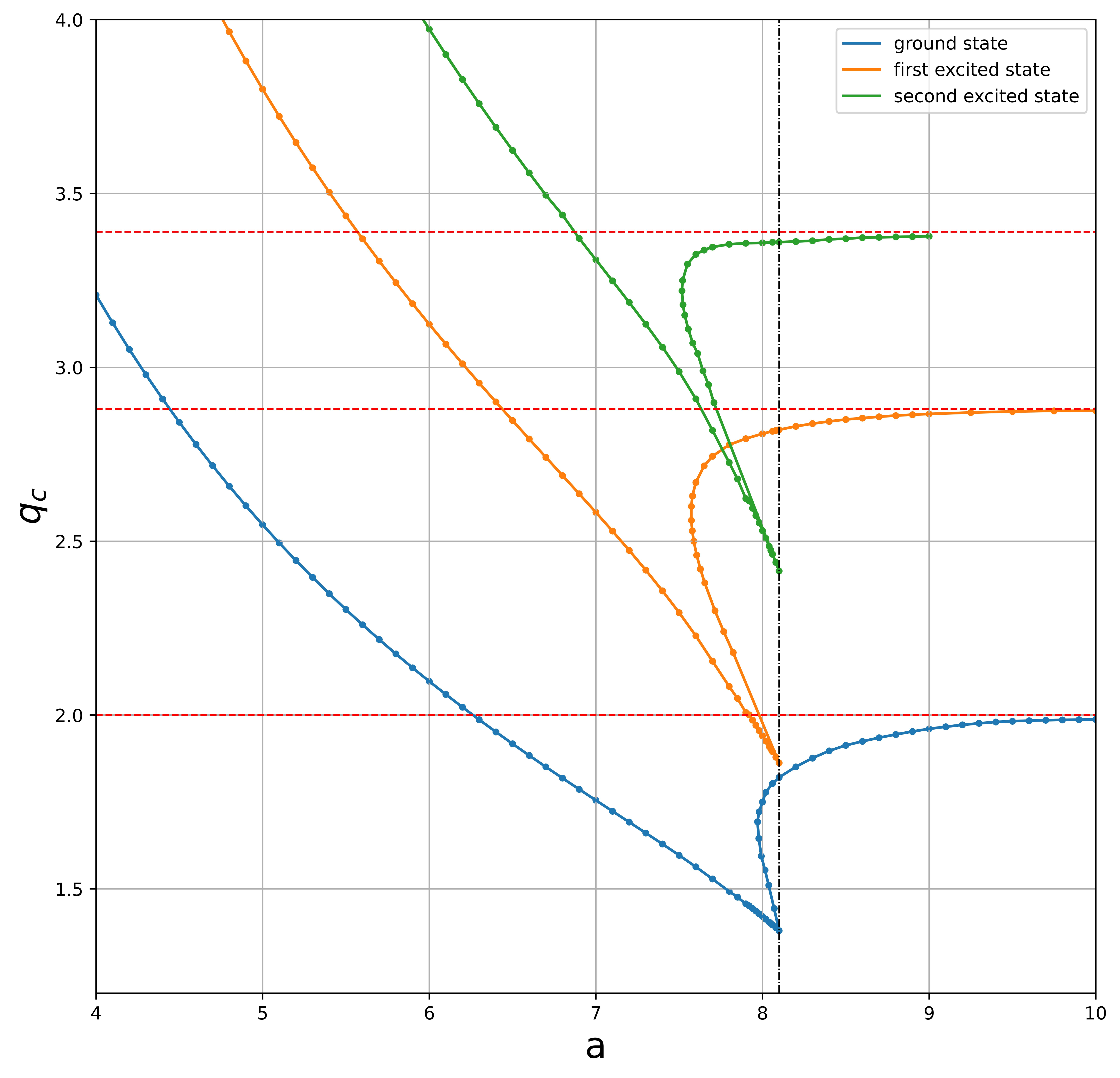}
\caption{The combined $q_c - a$ plot obtained by merging the $q_c^{singular} - a$ curves and old $q_c - a$ curves from the previous subsection, where the right graph shows a local magnification. The blue, yellow, and green lines respectively represent solutions for the ground state, first excited state, and second excited state. The horizontal red dashed lines indicate their respective asymptotic lines ($q_c^{bound} = 2$ for the ground state solution, $q_c^{bound} = 2.9$ for the first excited state solution, and $q_c^{bound} = 3.38$), while the black vertical dashed line represents $a = a_{max} = 8.08$.}
\label{fig:final}
\end{figure}

However, when we seek the critical charges corresponding to other values of $a$ and plot the corresponding $q_c^{singular} - a$ curves, we observe an asymptotic behavior for these $q_c^{singular}$. We obtained a new $q_c - a$ curve by combining the $q_c^{singular} - a$ curve we derived with the old $q_c - a$ curve obtained in the previous section, as shown in Fig.\ref{fig:final}. The blue, yellow, and green lines respectively represent the ground state, first excited state, and second excited state solutions. The red horizontal dashed line represents the asymptotic value $q_c^{bound}$ of $q_c$ ($q_c^{bound} = 2$ for the ground state solution \cite{Crisford:2017gsb}, $q_c^{bound} = 2.9$ for the first excited state solution, and $q_c^{bound} = 3.38$ for the second excited state solution). The black vertical dashed line represents $a = a_{max} = 8.08$ obtained from the previous subsection. As $a$ becomes very large, we find that the numerical solutions become highly sensitive to changes in $q$ around $q = q_c^{singular}$, making it difficult to obtain precise $q_c$ values for very large $a$. Furthermore, when extending from the ground state solution to the first excited state solution and then to the second excited state solution, this problem becomes even more severe. Hence, we only provide the $q_c^{singular} - a$ curve for the first excited state solution up to $a = 10$ and the $q_c^{singular} - a$ curve for the second excited state up to $a = 9$. However, even so, we can still clearly observe that the critical charges $q_c^{singular}$ corresponding to the appearance of the naked singularity exhibit a distinct asymptotic behavior. They seem to approach a bound $q_c^{bound}$, a bound that protects the Weak Cosmic Censorship Conjecture (WCCC)! Therefore, despite knowing very little about the results derived from the weak gravity conjecture (WGC) for excited state scalar fields, unable to provide a possible connection between WCCC and WGC as in \cite{Crisford:2017gsb}, we have still discovered a lower bound on the charge in excited state scalar field solutions that protects WCCC from being violated. By introducing a excited charged scalar field and requiring that the charge $q$ carried by the scalar field is greater than $q_c^{bound}$, we can repair the previous counterexamples that violate WCCC \cite{Horowitz:2016ezu,Crisford:2017zpi}, thus saving WCCC. 

\section{Summary}\label{sec4}
Inspired by the work \cite{Crisford:2017gsb}, we obtained numerical static solutions by coupling excited-state massive charged scalar fields with Einstein-Maxwell fields in four-dimensional anti-de Sitter spacetime. We realized that when treating the scalar field as a test field on a fixed spacetime background, the linear eigenequation derived from the scalar field equation's degeneration naturally encompass excited-state solutions. Subsequently, we solved the full nonlinear differential equations arising from coupling the scalar field equation with the background field equation and found corresponding solutions for the first and second excited states. We presented the spatial distribution of the excited-state scalar field and demonstrated the relationship between scalar condensate on the boundary $\left\langle\mathcal{O}_2\right\rangle$ and its variation with the boundary electric amplitude $a$. Compared with the ground-state scalar condensation which is obtained in \cite{Crisford:2017gsb}, we can distinctly observe the different configurations between the excited-state solutions and the ground-state solution.

We did not establish the stability analysis of these excited-state solutions. However, this doesn't hinder our discussion on whether a lower bound exists for the charges carried by the excited state scalar field to protect the Weak Cosmic Censorship Conjecture (WCCC). Our primary concern is whether WCCC relates to the basic requirements of matter fields in an unknown quantum gravity theory, and this doesn't depend on whether the matter fields are in a stable or unstable state. And we do have discovered a lower bound on the charge $q_c^{bound}$ for the excited-state scalar fields. As long as the charge carried by the scalar fields exceeds this threshold, regardless of how large the amplitude $a$ is, the Weak Cosmic Censorship Conjecture (WCCC) remains protected. This is remarkably similar to the results obtained for the ground state, even though the configurations of the excited-state and ground-state scalar fields are so different. This might imply that for excited-state scalar fields, the Weak Gravity Conjecture (WGC) could also offer a lower bound for the charge-to-mass ratio, potentially providing some modest assistance for future work related to WGC.

Our work serves as a valuable complement and extension to work \cite{Crisford:2017gsb}. However, the evolution of excited-state scalar fields and the potential existence of mixed states combining excited and ground states remain intriguing questions. In these scenarios, discussing the validity of the Weak Cosmic Censorship Conjecture (WCCC) would be a fascinating area of research. Such investigations would contribute to a better understanding of the essence of WCCC.

\section*{Acknowledgement}
We thank Shi-Xian Sun, Yang Song and Long-Xing Huang for helpful discussions. This work is supported by the National Key Research and Development Program of China
(Grant No. 2020YFC2201503) and the National Natural Science Foundation of China (Grant
No. 12047501 and No. 12275110).


\end{document}